\documentclass[twocolumn,showpacs]{revtex4}
\usepackage{amsmath,amssymb,dcolumn}
\usepackage{epsfig}

\newif\ifdviout
\dvioutfalse

\newcommand\rem[1]{}

\usepackage{amsmath}
\usepackage{amssymb}
\usepackage{mathrsfs}

\def\ifempty#1{\def\tmpdata{#1}\ifx\tmpdata\empty }

\def\linebreak{\hfill\break}

\def\bra<#1|{\langle #1\rvert}
\def\ket|#1>{\lvert#1 \rangle}
\def\braket<#1|#2>{\langle #1|#2 \rangle}

\def\otop#1{\hbox{$#1\kern-0.1em$\llap{\hbox{\raise1.7ex\hbox{$\scriptstyle\circ$}}}} }

\def\inpare#1{\left(#1\right)}
\def\bigpare(#1){\left(#1\right)}
\def\inrbra#1{\left\{ #1 \right\}}
\def\insbra#1{\left[ #1 \right]}

\def\bigbra[#1]{\left[ #1 \right]}

\def\therefore{\mbox{\setbox0=\hbox{X}\hbox{$\ldotp$}\raise0.7\ht0\hbox{$\ldotp$}\hbox{$\ldotp$}} \quad }
\def\because{\mbox{\setbox0=\hbox{X}\raise0.7\ht0\hbox{$\ldotp$}\hbox{$\ldotp$}\raise0.7\ht0\hbox{$\ldotp$}}\kern0pt }

\def\upin{\hbox{\setbox0=\hbox{$\cup$} \vrule width 0.05 \wd0 height \ht0 depth 0pt \kern - 0.5\wd0 \box0 }}
\def\Frac(#1/#2){\left(\frac{#1}{#2}\right)}

\def\Im{{\rm Im\,}}
\def\Re{{\rm Re\,}}

\def\sdprod{\mathrel{{\setbox0=\hbox{$\displaystyle\times$}\lower0.3\wd0\hbox{$\stackrel{\box0}{\scriptstyle\sim}$}}}}

\def\tosigma#1,{%
    \ifx\tmpindex\relax \def\tmpindex{#1} \let\next=\tosigma
    \else \ifnum\tmpindex=0 1 \else \sigma_\tmpindex \fi
          \ifx#1\relax  \let\next=\relax
          \else \otimes \let\next=\tosigma \def\tmpindex{#1} \fi
    \fi \next}
\def\tspb(#1){\let\tmpindex=\relax\tosigma#1,\relax,}
\def\Eqr#1{\begin{eqnarray} #1 \end{eqnarray}}

\def\Eqrsubl#1#2{\begin{subequations}
  \expandafter\ifx\csname Rlabel\endcsname \relax \label{#1}
  \else \Rlabel{#1} \fi \Eqr{#2}\end{subequations}}
\def\Bitm{\begin{itemize}}
\def\Eitm{\end{itemize}}
\def\Blist#1#2{\begin{list}{#1}{\parsep=0pt \itemsep=0pt%
  \listparindent=0pt #2}}
\def\Elist{\end{list}}
\long\def\ignore#1#2{\def\ignoreflag{#1}\long\def\tmptext{#2}
  \ifnum\ignoreflag>1 #2 \fi}

\def\imo{i}

\begin{document}

\title{Gravitational stability of simply rotating Myers-Perry black holes: tensorial perturbations}

\author{Hideo Kodama}\email{Hideo.Kodama@kek.jp}
\affiliation{Cosmophysics Group, IPNS, KEK and the Graduate University of Advanced
Studies, 1-1 Oho, Tsukuba 305-0801, Japan}

\author{R. A. Konoplya}\email{konoplya_roma@yahoo.com}
\affiliation{Department of Physics, Kyoto University, Kyoto 606-8501, Japan\\
Theoretical Astrophysics, Eberhard-Karls University of
T\"{u}bingen, T\"{u}bingen 72076, Germany}

\author{Alexander Zhidenko}\email{zhidenko@fma.if.usp.br}
\affiliation{Instituto de F\'{\i}sica, Universidade de S\~{a}o Paulo \\
C.P. 66318, 05315-970, S\~{a}o Paulo-SP, Brazil}

\begin{abstract}
We study the stability of $D \geq 7$ asymptotically flat black
holes rotating in a single two-plane against tensor-type
gravitational perturbations. The extensive search of quasinormal
modes for these black holes did not indicate any presence of
growing modes, implying the stability of simply rotating
Myers-Perry black holes against tensor-type perturbations.
\end{abstract}

\maketitle

\section{Introduction}

Higher-dimensional gravity appears in the landscape of modern
physics as a result of the string theory or in various brane-world
models. Both, string theory and brane-world contexts imply the
existence of higher dimensional black holes. For the four
dimensional gravity, there exists a uniqueness theorem that states
that the asymptotically flat stationary black hole is described by
the well-known Kerr solution. In higher dimensions, such a
uniqueness theorem does not hold any more, but a great variety of
solutions describing black objects, such as black holes, strings,
branes, rings, saturns, exist instead
\cite{Reall-review}. Therefore, the stability of rotating black
holes in higher than four space-time dimensions is an open
appealing problem
\cite{Kodama:2007ph}, \cite{Kodama.H2008}. Stability could help to
select among the variety of "black" solutions those which could be
realized in nature. Another motivation to study the evolution of
higher-dimensional black hole perturbations comes from the possibility to observe mini-black holes in future experiments
with cosmic showers or in the Large Hadron Collider
\cite{Kanti-review}.

The stability of spherically symmetric static black holes has been
intensively studied in recent few years. It started from the
papers by Ishibashi and Kodama \cite{Kodama1},
\cite{Kodama2}, \cite{Kodama3}, who, using the gauge-invariant
formalism \cite{Kodama.H&Ishibashi&Seto2000}, proved that the
D-dimensional generalization of the Schwarzschild black hole,
described by the Tangherlini metric, is stable. Then, the
stability of higher-dimensional black holes allowing for a
cosmological constant was shown in \cite{KonoplyaNPB} for
$D=5,6,\ldots11$ Schwarzschild-de Sitter (SdS) and in
\cite{Konoplya:2008rq} for $D=5,6,\ldots11$
Reissner-Nordstr\"om-anti-de Sitter (RNAdS) black holes. The
D-dimensional Reissner-Nordstr\"om black hole was shown to be
stable, although Reissner-Nordstr\"om-de Sitter black holes proved
to be unstable in higher than six space-time dimensions
\cite{Konoplya:2008au}, when both the charge and $\Lambda$- term
are large enough. The stability of black holes in higher
dimensional generalizations of the Einstein gravity, such as the
Gauss-Bonnet and Lovelock theories, was considered in
\cite{Takahashi:2009dz},
\cite{Beroiz:2007gp}, \cite{Konoplya:2008ix2}, and an instability
for large values of the Gauss-Bonnet coupling $\alpha$ was found.
This Gauss-Bonnet instability was shown to have quite peculiar
behavior: the instability starts dominating after a long period of
damped quasinormal oscillations. In addition, the stability
analysis and time domain evolution of perturbations were
considered for the Kaluza-Klein type solutions, such as (squashed)
Kaluza-Klein black holes \cite{Ishihara:2008re} or black strings
\cite{Konoplya:2008yy}.

While the stability of static black holes in higher dimensions is
relatively well studied, the stability of rotating black holes for
$D>4$ is still an open question, because the separation of
variables for all types of perturbations is a complicated problem.
Nevertheless, some particular results have been obtained for the
Myers-Perry black holes with all equal angular momenta, though
only for either tensor-type perturbations \cite{Kunduri:2006qa} or
for the so-called zero mode \cite{Murata:2008yx}.

In our previous paper \cite{Kodama4}, we have started the
investigation of gravitational perturbations of a much more
interesting type of higher-dimensional rotating black holes,
simply rotating black holes. These black holes have only one
non-vanishing angular momentum component and are rotating in a
single two-plane. Thus, they can be interpreted in brane-world
scenarios as black holes created in particle collisions occurred
on our brane. The perturbation equations for the $D>6$ dimensional
simply rotating black holes, allowing for a cosmological constant,
were reduced in
\cite{Kodama4} to the wave-like form, and gravitational spectrum
of asymptotically-AdS solutions was considered. The asymptotically
AdS rotating black holes proved to be gravitationally unstable on
superradiant modes only, and with a tiny instability growth rate
\cite{Kodama4}.

In the present paper we shall continue research that started in
\cite{Kodama4} and consider tensor-type gravitational spectrum of
asymptotically flat, Myers-Perry black holes. The quasinormal modes beside their importance for stability
proof, may have useful implications for the study of the evolution of
mini higher-dimensional black holes, because quasinormal
modes dominate at late time of the bolding phase
\cite{Kanti-review} of an evaporating black hole.

Using the Frobenius method for the angular part of the
perturbation equations and both WKB and Frobenius methods for the
radial part, we have found the quasinormal modes of
simply-rotating Myers-Perry black holes. The paper is organized as
follows: Sec. II gives basic information on simply-rotating MP
black holes. Sec. III is devoted to description of the numerical
methods that we used for finding the quasinormal modes. Sec. IV
discusses the obtained numerical results.

\begin{widetext}
\section{Basic formulas}

The metric for the higher-dimensional Myers-Perry black hole can be written in the Boyer-Lindquist coordinates as
\Eqr{
ds^2 &=& \frac{1}{\rho^2}\insbra{\frac{2M}{r^{n-1}} -\rho^2} dt^2
   -\frac{4aM\sin^2\theta}{\rho^2 r^{n-1}}dtd\phi
+\frac{\sin^2\theta}{\rho^2}\insbra{(r^2+a^2)\rho^2+\frac{2a^2M}{r^{n-1}}\sin^2\theta}
   d\phi^2
  \notag\\&&
   + \frac{\rho^2}{\Delta}dr^2+ \rho^2d\theta^2
   +r^2\cos^2\theta d\Omega_n^2,
\label{KadS:metric:BL}
}
where $n = D-4$, $d\Omega_n^2$ is the metric of the $n$-dimensional unit sphere, and
\Eqr{
\Delta := (r^2+a^2)- \frac{2M}{r^{n-1}},
&\qquad& \rho^2:= r^2 + a^2\cos^2\theta.
}
Here $M$ is the mass of the black hole and $a$ is the angular momentum per unit mass.

The equations for the tensor-type gravitational perturbations of this black hole are reduced to the coupled equations \cite{Kodama4} for the angular part
\begin{eqnarray}\label{angular-part}
2(1-x^2)S''(x)+(-1 + n - (3 + n) x) S'(x)
+\left(\frac{\mu}{2}+\frac{a^2\omega^2(x-1)}{4}+\frac{m^2}{x-1}-\frac{\ell (\ell + n-1)}{1 + x}\right)S(x)&=&0,
\end{eqnarray}
$$x=\cos(2\theta), \qquad n=3,4,5\ldots, \qquad \ell=2,3,4\ldots, \qquad m=0,\pm1,\pm2\ldots,$$
and for the radial part
\begin{equation}\label{wave-like-flat}
\frac{d^2 P(r)}{dr_\star^2}
+Q(r)P(r)=0,
\end{equation}
\begin{eqnarray}\label{WKBpot} Q(r)&=&\left(\omega-\frac{2Mam}{(r^2+a^2)^2r^{n-1}}\right)^2-\frac{\Delta(r)}{(r^2+a^2)^2}\left(\mu -\frac{a^2 m^2}{(r^2+a^2)^2} \inrbra{r^2+a^2+ \frac{2M}{r^{n-1}} } + \frac{n(n+2)}{4}+\right.\\\nonumber&&\left.+\inpare{l+\frac{n}{2}}\inpare{l+\frac{n}{2}-1} \frac{a^2}{r^2} +\frac{a^2}{r^2+a^2}++\frac{\inrbra{(n+2)r^2+na^2}^2-8a^2 r^2}{2(r^2+a^2)^2}
      \frac{M}{r^{n+1}}\right).
\end{eqnarray}
with respect to the frequency $\omega$ and the separation constant
$\mu$. Here, we have defined the tortoise coordinate $r_\star$ as
$$
dr_\star=\frac{r^2+a^2}{\Delta(r)}dr.
$$
\end{widetext}

\section{Numerical method for finding of the quasinormal spectrum}

\subsection{The angular part}\label{sec:angular-part}

The differential equation for the angular part
(\ref{angular-part}) can be solved numerically in the same way as
it was done in \cite{Suzuki:1998vy}. Namely, the second-order
differential equation (\ref{angular-part}) is reduced to an
algebraic equation with an infinite continued fraction
\cite{Kodama4}. This equation can be solved numerically to determine the separation constant $\mu$ for each value of
$\omega$. Thus, we can find numerically the function
$\mu_j(\omega)$, where $j=0,1,2\ldots$ is an integer number.

The number $j$ enumerates the eigenvalues of $\mu$. If the black
hole does not rotate ($a=0$), the equation for the angular part
can be solved exactly to yield \cite{Kodama4}
\begin{equation}\label{angular-part-nonrotating}
\mu = (2j+\ell+|m|)(\ell+2j+|m|+n+1),
\end{equation}
which coincides with the eigenvalue for the harmonic function on
the unit $(n+2)$-sphere, with the multi-pole number
$(2j+\ell+|m|)=2,3,4\ldots$.

If the rotation parameter $a$ is non-vanishing, the eigenvalues
$\mu$ are non-integer and complex. For any value of $\omega$, we
can enumerate them by the non-negative integer $j$ in the
following way.
\begin{enumerate}
\item We start from the non-rotating black hole and find exactly the value of $\mu$ for the corresponding $j$.
\item We increase the rotation parameter by a very small value and search for the closest to the previously found solution for $\mu$.
\item We repeat the previous step until any required value of $a$ is reached.
\end{enumerate}
In this way we are able to enumerate the values of the separation
constant $\mu$ for any $\omega$ and $a$. In order to decrease the
computation time, in practice, when searching for quasi-normal
modes, we start from the known values of the quasi-normal
frequencies for a non-rotating black hole
\cite{Konoplya:2003dd} and, increasing step by step the value of
$a$, find the closest pair of $\mu$ and $\omega$ to the previously
found one. An alternative approach for the computation of the
angular separation constant has been recently suggested in
\cite{Cho:2009wf}.

\subsection{The radial part}

In order to simplify our notations, we will parameterise the black
hole mass by the black hole horizon radius $r_+$:
\begin{equation}
2M=r_+^{n-1}(r_+^2+a^2).
\end{equation}
In solving the radial part of the perturbation equation, we use
two alternative approaches: the numerical Frobenius method
\cite{Leaver:1985ax} and the semi-analytical JWKB approximation
\cite{WKB}.

Let us start from the description of the Frobenius method. We are
searching for the solution of the equation (\ref{wave-like-flat})
with the quasi-normal boundary conditions that requires pure
in-coming waves at the event horizon and pure out-going waves at
spatial infinity. For a black hole in the asymptotically flat
background, they imply a purely ingoing wave at the event horizon
$r_+$ and purely outgoing wave at the spatial infinity. If $P(r)$
satisfies the quasi-normal boundary conditions, then it can be
written in the following form

\begin{widetext}
\begin{equation}\label{singularities_expand}
P(r)=\left(1-\frac{r_+}{r}\right)^{\displaystyle-\imo\left(\omega -
   \frac{2 M a m}{(r_+^2 + a^2)^2 r_+^{n - 1}}\right) \frac{r_+^2 + a^2}{\Delta'(r_+)}}\times\frac{\sqrt{r^2+a^2}}{r}\times e^{\displaystyle\imo\omega r}\times F(r),
\end{equation}
\end{widetext}
where $F(r)$ is a regular function at the event horizon $r=r_+$ and at the spatial infinity $r=\infty$.

In order to expand $F(r)$ in the convergent series in the region outside the event horizon, we choose the variable
$$z=\frac{r-r_+}{r-r_i},$$
where $r_i<r_+$ is chosen so that all the singular points of the equation (\ref{wave-like-flat}), except the event horizon and the spatial infinity, are located outside the unit circle $|z|>1$. These singular points satisfy the equation
$$\Delta(r)=0.$$
Fortunately, the structure of the singular points allows us to fix
$r_i$ for a physically reasonable range of parameters.

After the value of $r_i$ is fixed, we expand the function $F$ into the series
\begin{equation}\label{Frobenius_series}
F(z)=\sum_{k=0}^\infty b_k z^k.
\end{equation}
Substituting (\ref{Frobenius_series}) and
(\ref{singularities_expand}) into (\ref{wave-like-flat}), we find
the recurrence relation for the series coefficients $b_i$.
Following \cite{Zhidenko:2006rs}, we are able to find the equation
with an infinite continued fraction with respect to $\omega$,
which can be solved numerically.

The Frobenius method, being based on a convergent procedure,
allows us to find quasinormal modes (QNMs) with any required
accuracy, at least theoretically. Practically, for some range of
parameters (in particularly for large azimuthal numbers) the
convergence is too slow and the calculation cannot be easily done
by a personal computer. In the range of parameters of slow
Frobenius convergence, we use the WKB approximation
\cite{WKB}. The WKB formula (\ref{WKBformula}) was effectively
used in a lot of papers
\cite{WKBuse}, and here the WKB accuracy will be checked with the
help of the the Frobenius method. The 6-th order WKB formula reads
\begin{equation}\label{WKBformula}
\frac{\imath Q_{0}}{\sqrt{2 Q_{0}''}} - \sum_{i=2}^{i=6} \Lambda_{i} = N+\frac{1}{2},\qquad N=0,1,2\ldots,
\end{equation}
where the correction terms $\Lambda_{i}$ were obtained in
\cite{WKB} and \cite{WKBorder}. Here $Q_{0}$ and $Q_{0}''$ denote values of $Q$ and its second derivative at its maximum with respect to the tortoise
coordinate $r_\star$, respectively. Since the function
$\mu(\omega)$ is found only numerically, in order to find the
roots of (\ref{WKBformula}), we use the following algorithm:
\begin{enumerate}
\item For a given value of $\omega$, we find the value of $\mu$ using the technique described in the section \ref{sec:angular-part}.
\item We substitute the numerical values of $\omega$ and $\mu$ into (\ref{WKBpot}) and find $Q$ as a complex analytical function of $r$.
\item The analytical continuation of the WKB formula into the complex plane implies that ``maximum'' of the potential peak is a complex value, which must be correctly chosen among the extrema of the potential $Q$. Since the WKB formula provides a sufficient accuracy only when $Re(\omega)\gtrsim Im(\omega)$, we will not consider the case, when the imaginary part of the quasi-normal frequency is large. Fortunately, when the imaginary part of $\omega$ is small, the imaginary part of $\mu$ is just a correction to its real part. Therefore, the real part of $Q(r)$ is much larger than its imaginary part, and the analytical continuation of the potential peak is trivial. We choose the location $r_{ext}$ of the extremum of $Q(r)$, with a real part that is larger than the value of the event horizon and, at the same time, has the smallest imaginary part. This choice of the potential peak does not lead to any ambiguities for the considered parametric range, is close to the real potential peak for the non-rotating case and its searching can be easily programmed.
\item After the potential peak is fixed, we compare the frequency $\omega_{WKB}$ found with the formula (\ref{WKBformula}) with our initial guess $\omega$.
\end{enumerate}
We search for the numerical minimum of the absolute value of the
difference between the initial guess and the WKB formula
$|\omega_{WKB}(\omega)-\omega|$, which is a function that can be
calculated numerically using the above technique.

\begin{figure*}
\includegraphics[width=.45\textwidth]{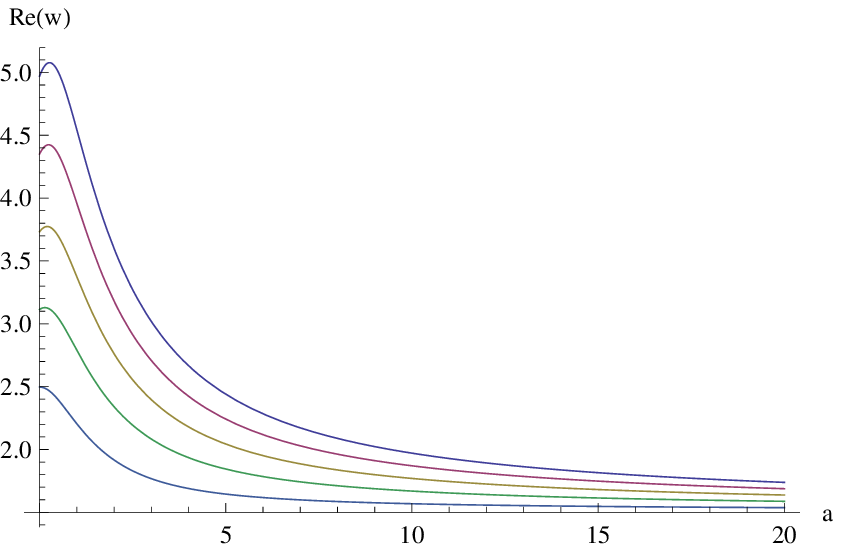}
\includegraphics[width=.45\textwidth]{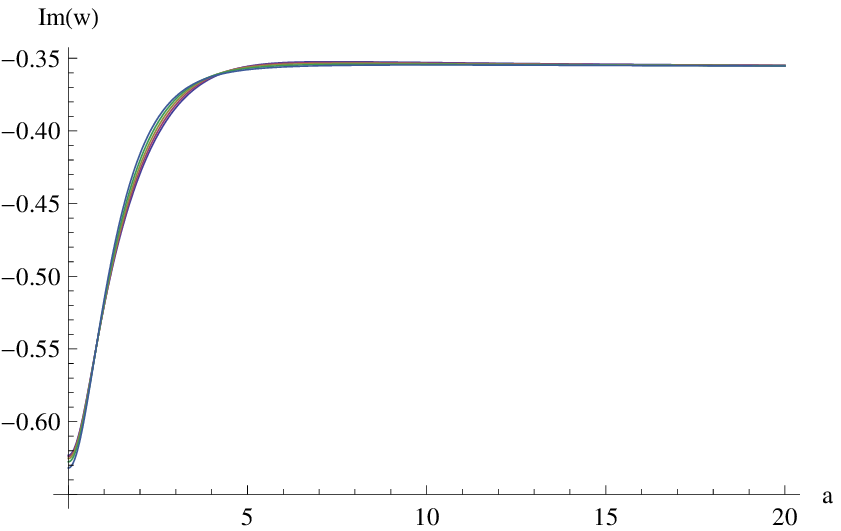}
\caption{\label{fig:D7_apm}  Quasinormal modes as a function of $a$ obtained by the Frobenius method for $D=7$, $l=2$, $j=0$: $m=4$ (blue), $m=3$ (red), $m=2$ (yellow), $m=1$ (green), $m=0$ (light blue). Higher values of $m$ correspond to larger real part of $\omega$.}
\end{figure*}

\begin{figure*}
\includegraphics[width=.45\textwidth]{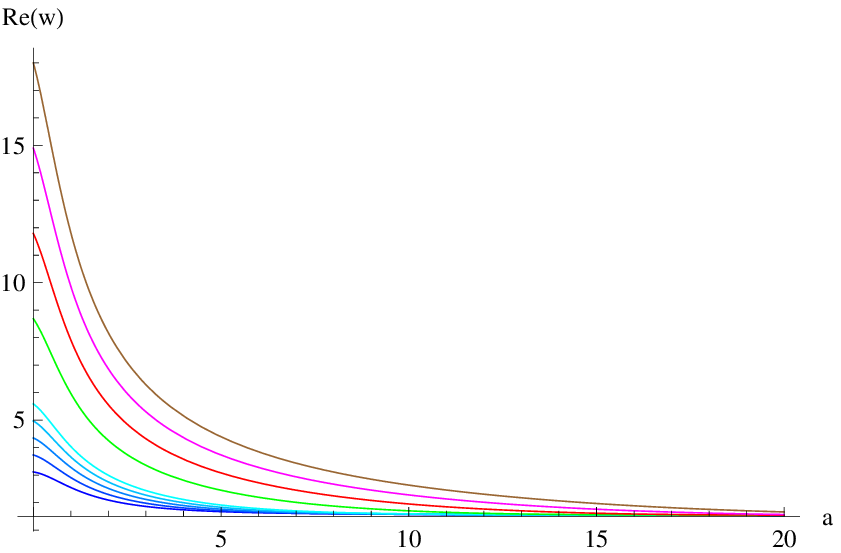}
\includegraphics[width=.45\textwidth]{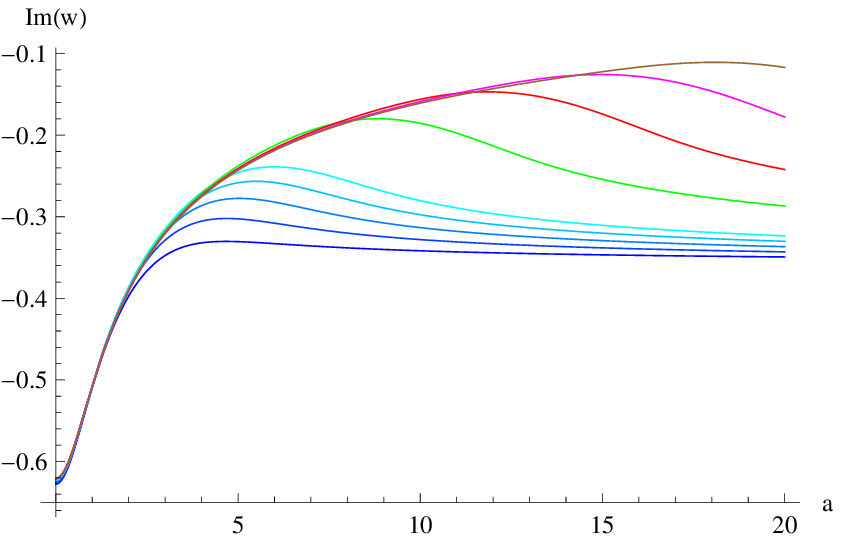}
\caption{\label{fig:D7_anm}  Quasinormal modes obtained by the Frobenius method for $D=7$, $l=2$, $j=0$: $m=-1$ (blue), $m=-2$, $m=-3$, $m=-4$, $m=-5$ (light blue), $m=-10$ (green), $m=-15$ (red), $m=-20$ (magenta), $m=-25$ (brown). Higher negative values of $m$ correspond to the larger real and imaginary part of $\omega$. The imaginary part of the QN frequency stays negative, implying stability against perturbations with the high negative azimuthal number $m$.}
\end{figure*}

\begin{figure*}
\includegraphics[width=.45\textwidth]{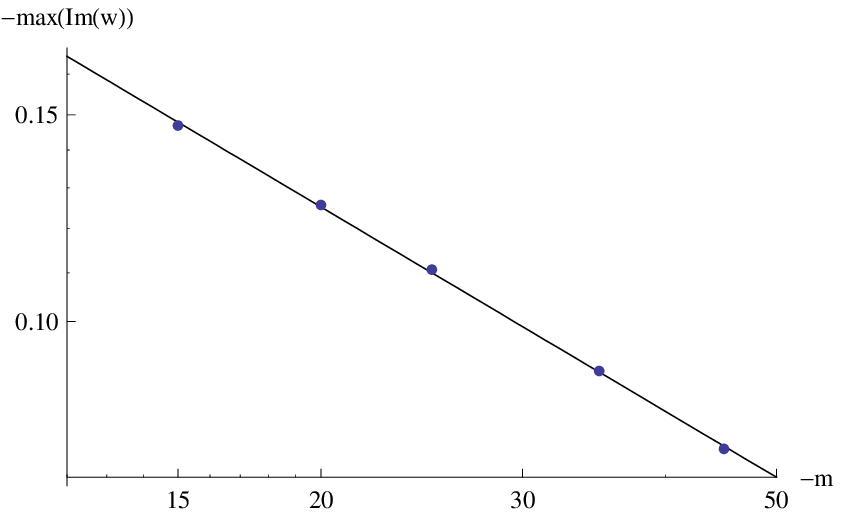}
\includegraphics[width=.45\textwidth]{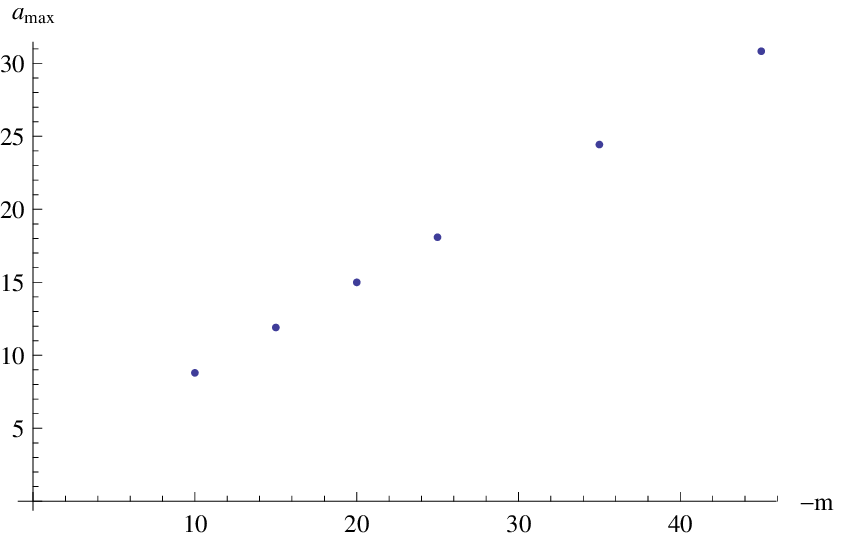}
\caption{\label{fig:D7_fit} Maximum value of the imaginary part of the quasi-normal mode for negative values of $m$ together with the found fit $max(Im(\omega)\propto (-m)^{-\alpha}~(\alpha>0)$ for $D=7$, $l=2$, $j=0$ (left figure, log-log scale) and corresponding values of the rotation parameter $a_{max}$ as a function of $-m$ (right figure).}
\end{figure*}

\begin{figure*}
\includegraphics[width=.45\textwidth]{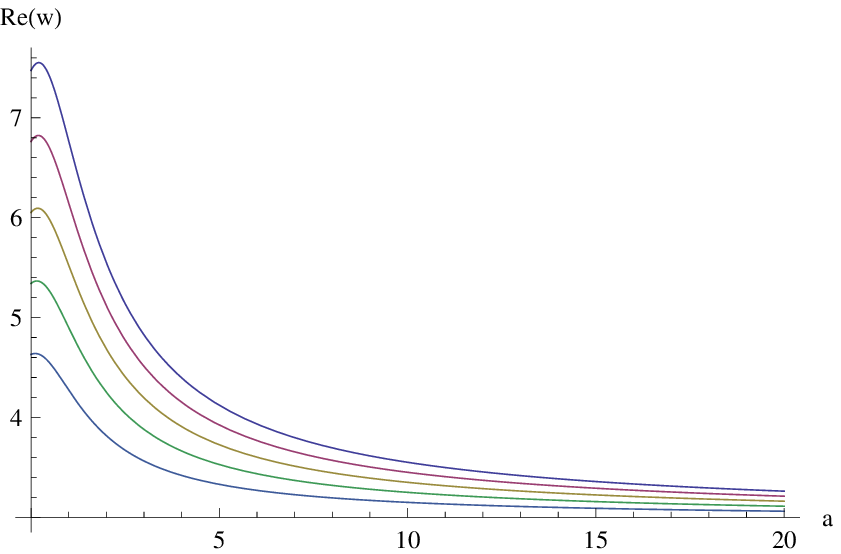}
\includegraphics[width=.45\textwidth]{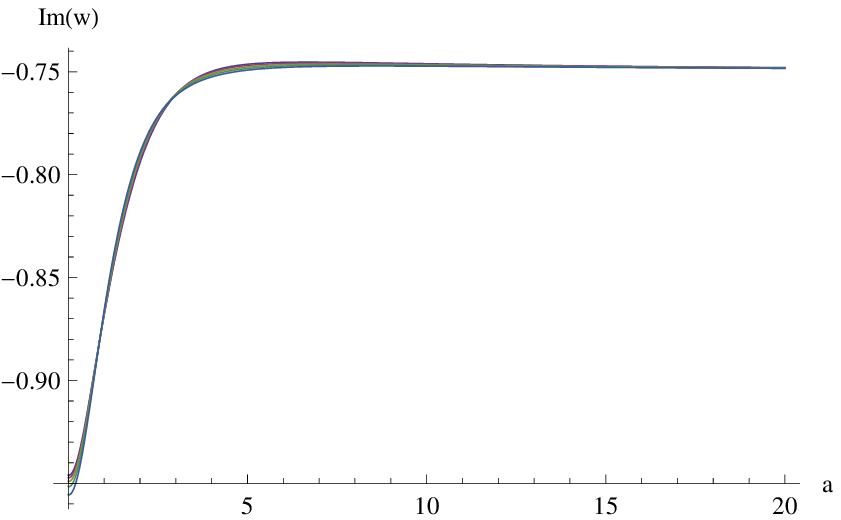}
\caption{\label{fig:D11_a} Quasinormal modes obtained by the Frobenius method for $D=10$, $l=2$, $j=0$: $m=5$ (blue), $m=4$ (red), $m=3$ (yellow), $m=2$ (green), $m=1$ (light blue). Higher values of $l$ correspond to larger real part of $\omega$.}
\end{figure*}

\begin{figure*}
\includegraphics[width=.45\textwidth]{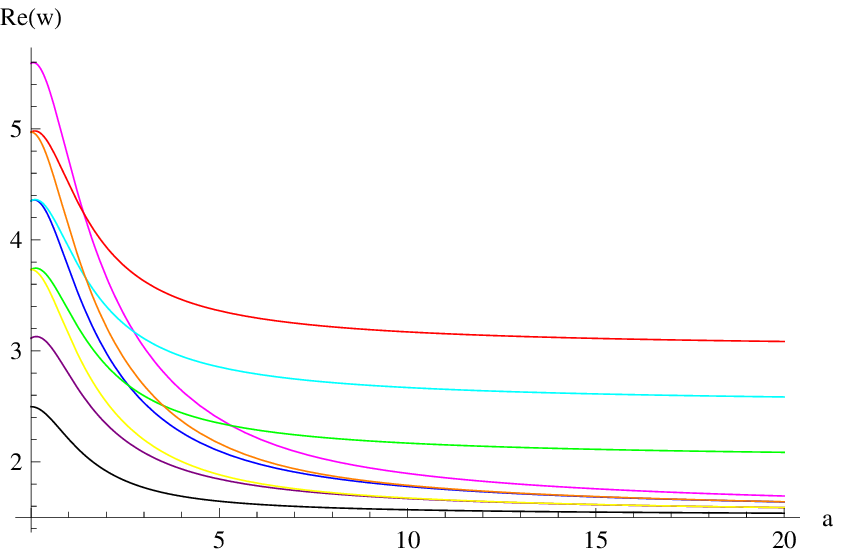}
\includegraphics[width=.45\textwidth]{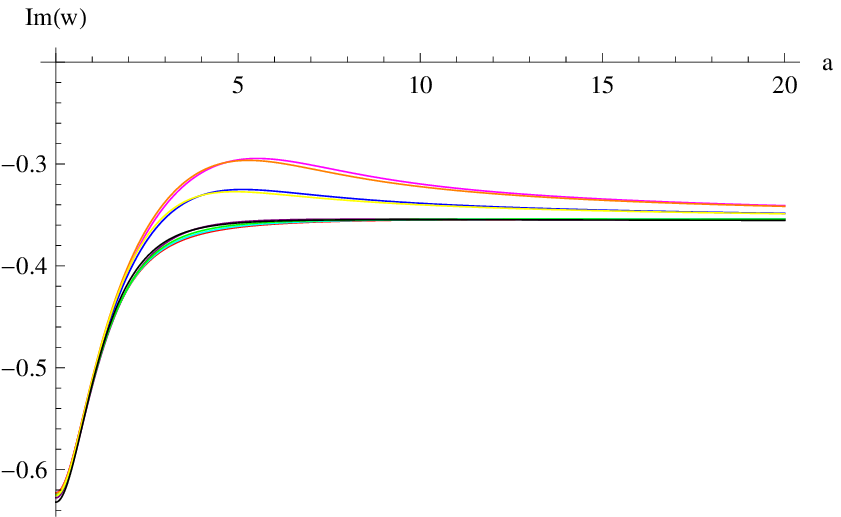}
\caption{\label{fig:D7a} Quasinormal modes obtained by the Frobenius method for $D=7$: ($l=2$, $m=0$, $j=0$ - black), ($l=2$, $m=0$, $j=1$ - yellow), ($l=2$, $m=0$, $j=2$ - orange), ($l=2$, $m=1$, $j=0$ - purple), ($l=2$, $m=1$, $j=1$ - blue), ($l=2$, $m=1$, $j=2$ - magenta), ($l=3$, $m=1$, $j=0$ - green), ($l=4$, $m=1$, $j=0$ - cyan), ($l=5$, $m=1$, $j=0$ - red). The larger values of $j$ correspond to the smaller values of the imaginary part  of $\omega$. The real part increases with $l+m+2j$ for small rotation. When $a$ is large, the real part for the multipole index $j\neq0$ approaches zero, while that for $j=0$ approaches a constant that increases with  $l$.}
\end{figure*}

\begin{figure*}
\includegraphics[width=.45\textwidth]{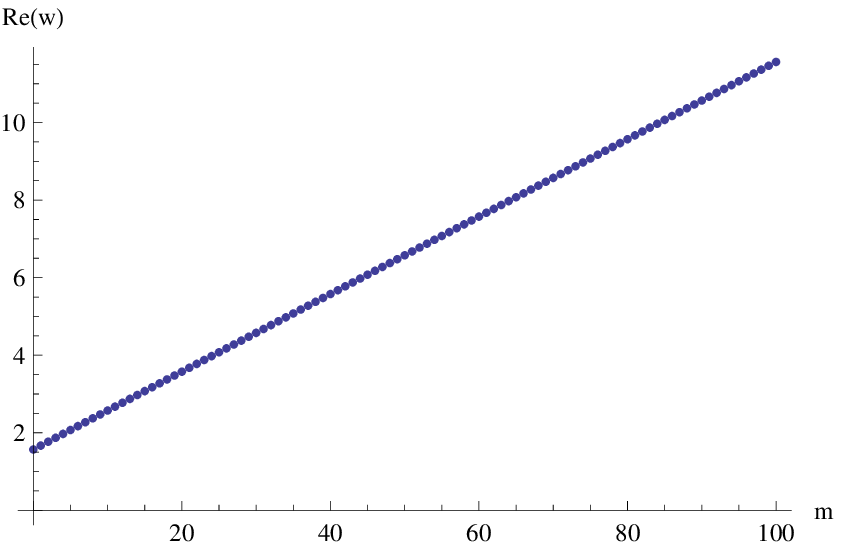}
\includegraphics[width=.45\textwidth]{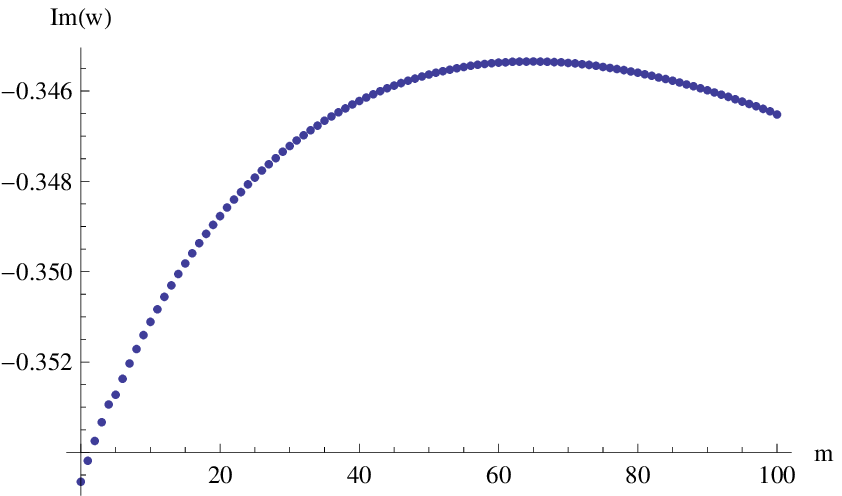}
\caption{\label{fig:D7a10} Dependence of quasinormal modes obtained by the WKB method on the azimuthal number $m$ for $D=7$, $l=2$, $j=0$, $a=10$.}
\end{figure*}

\begin{figure*}
\includegraphics[width=.45\textwidth]{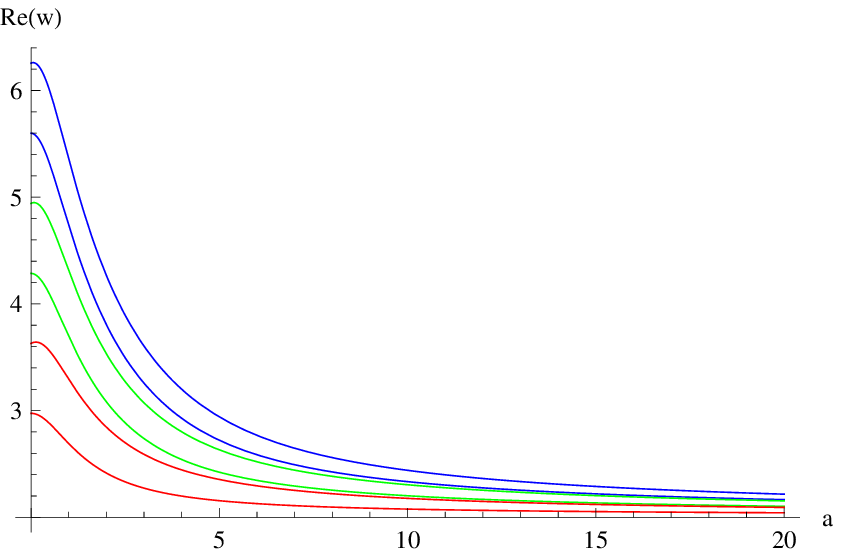}
\includegraphics[width=.45\textwidth]{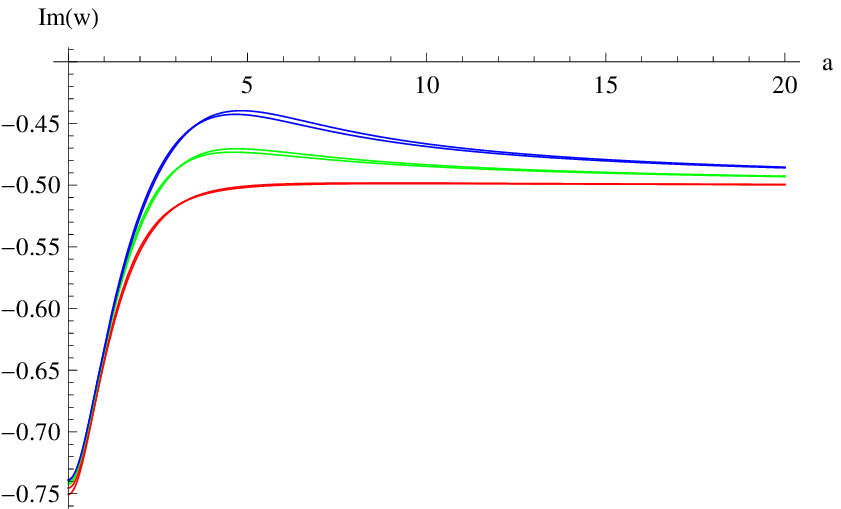}
\caption{\label{fig:D8} Quasinormal modes as a function of $a$ for $D=8$, $l=2$, $m=0,1$: $j=0$ (red), $j=1$ (green), $j=2$ (blue). Higher $m$ and $j$ have larger real and imaginary part of $\omega$}
\end{figure*}
\begin{figure*}
\includegraphics[width=.45\textwidth]{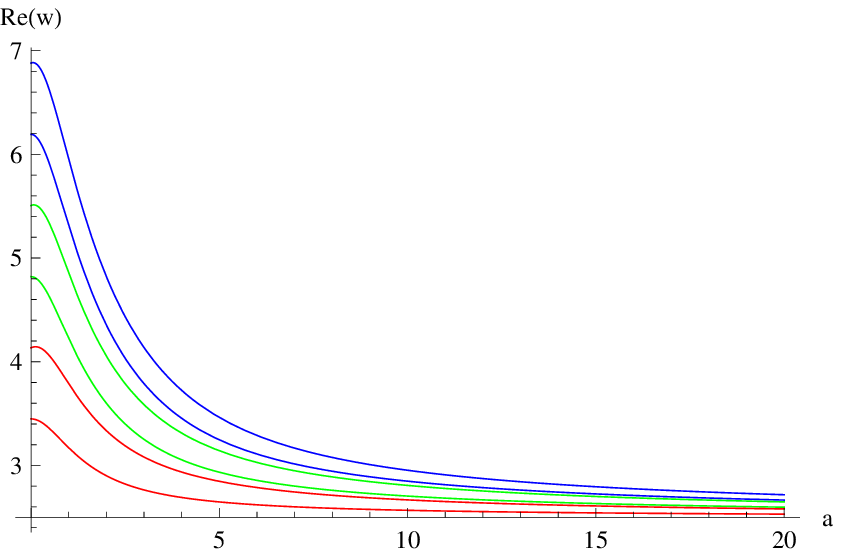}
\includegraphics[width=.45\textwidth]{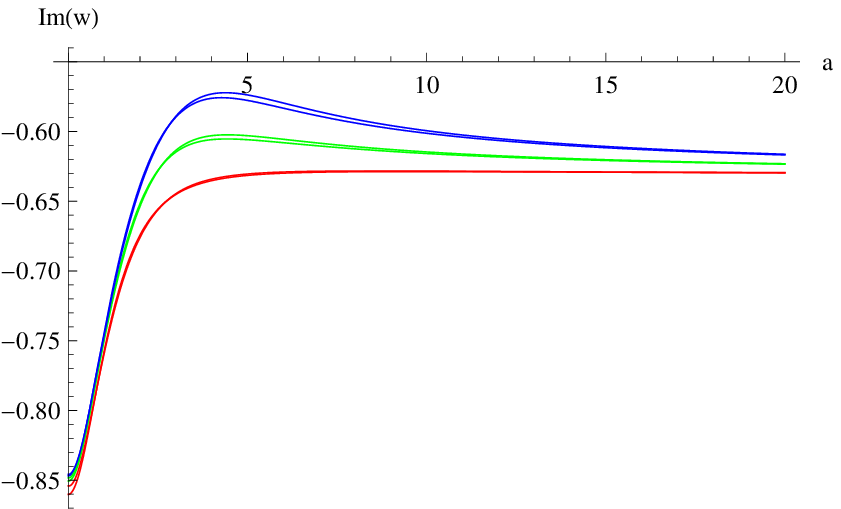}
\caption{\label{fig:D9} Quasinormal modes as a function of $a$ for $D=9$, $l=2$, $m=0,1$: $j=0$ (red), $j=1$ (green), $j=2$ (blue). Higher $m$ and $j$ have larger real and imaginary part of $\omega$.}
\end{figure*}

\begin{figure*}
\includegraphics[width=.45\textwidth]{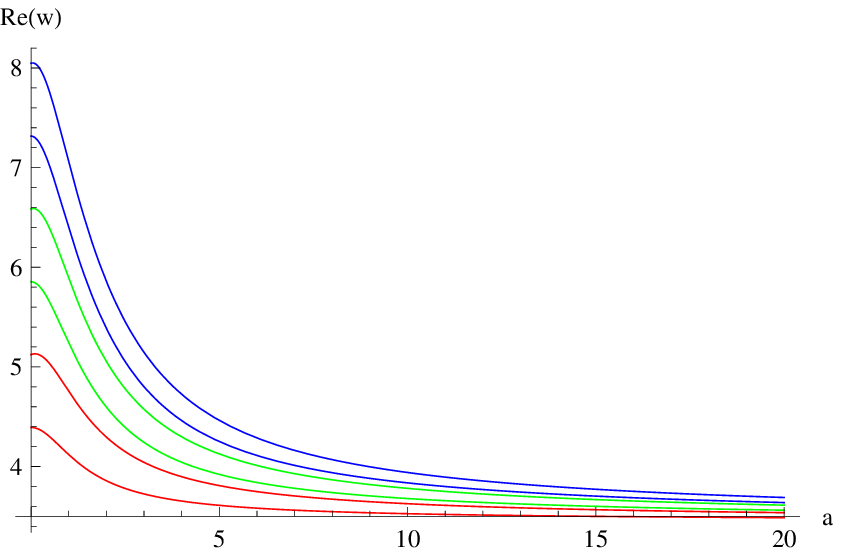}
\includegraphics[width=.45\textwidth]{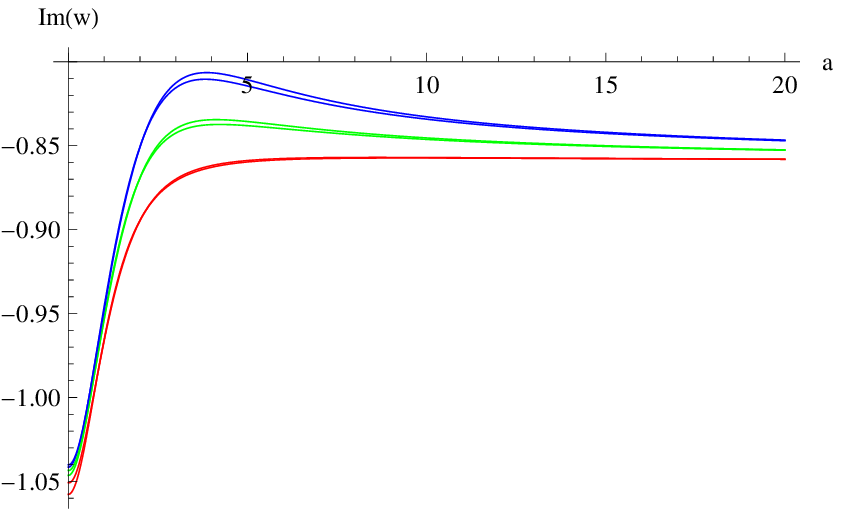}
\caption{\label{fig:D11} Quasinormal modes as a function of $a$ for $D=11$, $l=2$, $m=0,1$: $j=0$ (red), $j=1$ (green), $j=2$ (blue). Higher $m$ and $j$ have larger real and imaginary part of $\omega$.}
\end{figure*}

\begin{figure*}
\includegraphics[width=.45\textwidth]{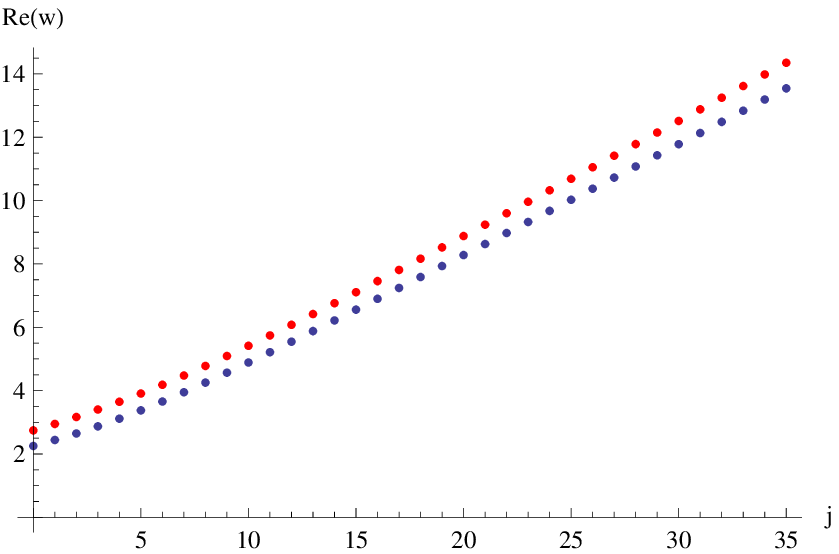}
\includegraphics[width=.45\textwidth]{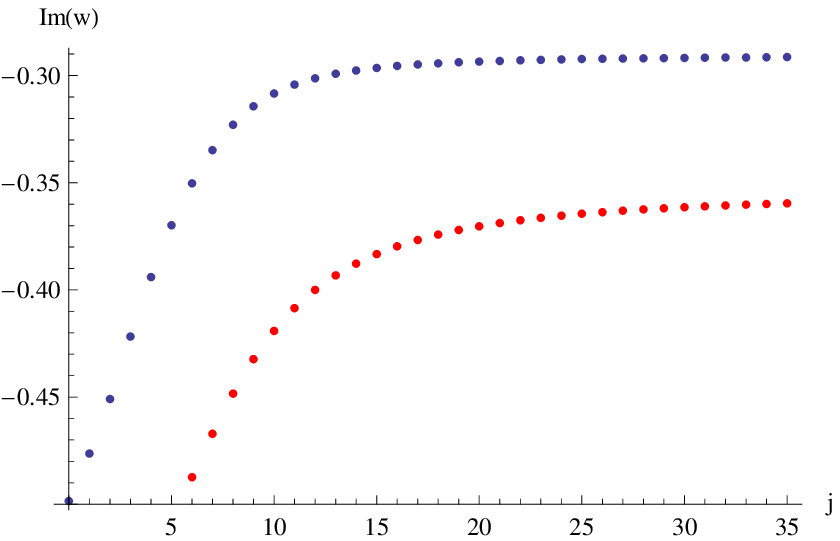}
\caption{\label{fig:a7m1} Dependence of the quasinormal modes on $j$ for $D=8$ (blue) and $D=9$ (red), $a=7$, $m=1$. The plots with larger $Re(\omega)$ and $Im(\omega)$ correspond to the larger $D$.}
\end{figure*}

\section{Quasinormal modes and stability}

When we write the quasinormal frequency $\omega$ as
\begin{equation}
\omega = \Re (\omega) + i \Im (\omega),
\end{equation}
$\Im (\omega) > 0$ corresponds to an unstable (growing) mode, and
$\Im (\omega) < 0$ corresponds to a stable (damped) mode.

The quasinormal modes of a higher-dimensional simply rotating
black hole are labeled by the quantum numbers $\ell$, $m$ and $j$,
and also depend on the black hole parameters $M$ and $a$ as well
as on the number of space-time dimensions $D$. Thus, the
dependence of the spectrum on the above six parameters must be
carefully considered in order to discard or observe instability.
We have limited the calculations to $D=7, 8, 9, 10$ and $11$.
Keeping the event horizon radius $r_+$ fixed, we can measure the
QN frequencies in units $r_{+}^{-1}$. Thus, the dependence of the
quasinormal behavior on the four essential parameters must be
studied. As all of them $\ell$, $m$, $j$, $a$ are not limited, in
principle, we have to check the QN frequency as a function of each
parameter starting from its minimal value and until reaching
\emph{the asymptotic region} where the frequency function does not
change its behavior qualitatively. However, this task would require us to show here hundreds of
plots and would take enormous computer time. Therefore, we were obliged to adopt the
following practical strategy. If the frequency
$\omega$ as a function of some parameter does not show any
tendency to instability and the lowest (longest lived) modes stay
very far from the instability region, we stopped the calculation
at some reasonable value of the parameter. Strictly speaking, this
does not allow us to establish mathematically rigorous proof of
stability, but rather only give a numerical evidence of stability. In presenting the obtained results, in order to reduce the number
of plots, we selected most representative ones with qualitatively
different behavior. If plots are similar (for instance for various
$D$ and the same other parameters) we have shown only one representative plot of the
whole ``class''.

Comparison of the accurate data obtained by the Frobenius  method
with the WKB values shows that the WKB approximation becomes better
as the azimuthal number $m$ becomes larger, 
because it corresponds to larger values of the real part of the QN frequency. Indeed, the WKB approximation
works when $|Re(\omega)|
\gg |Im(\omega)|$. Thus, the limit of perfect WKB accuracy, which corresponds to the
geometric optics approximation, should be achieved at $m
\rightarrow \infty$.

Figures \ref{fig:D7_apm}, \ref{fig:D7_anm} and
\ref{fig:D11_a} show the dependence of the quasinormal
frequency, obtained by the Frobenius method, on the rotation
parameter a for a few low values of m and the lowest values of $l$
and $j$. QNMs for a few low values of $l$ and $j$ can be found in
Figure \ref{fig:D7a}. There, one can see that $Re(\omega)$ almost
monotonically decreases (except for a small peak at very small
$a$), when $a$ is increased, while $Im(\omega)$ is growing until
reaching its maximum, and then falling off to some asymptotic
constant value. As the absolute value of $Im(\omega)$ decreases
when $m$ increases, it is important to find QNMs for sufficiently
large $m$, in order to rule out the instability.  On figures Fig.
\ref{fig:D7a10} one can see the high $m$ behavior for a fixed $a$
and $D$. Apparently $Re(\omega)$ is equidistant in the limit $m
\rightarrow \infty$. The absolute value of $Im(\omega)$ approaches
some maximum and then starts to fall off at large $m$, reaching
some constant value asymptotically (see Fig.
\ref{fig:D7a10}).

In the figure \ref{fig:D7_anm} one can see that for large negative
$m$ the maximum of the imaginary part is higher (closer to zero)
and occurs for larger value of the rotation parameter $a$. One
might ask himself if there is sufficiently high value of $m$ for
which the imaginary part crosses zero at some point leading to the
instability for larger values of $a$. Our calculations show that
this does not take place. In the left figure \ref{fig:D7_fit} we
plot the dependence of the maximum value of $\Im(\omega)$ as a
function of $-m$ in the logarithmic-logarithmic scale. We see that
this value is well approximated by the potential law
$max(\Im(\omega))\propto(-m)^{-\alpha}~(\alpha>0)$. If this is
valid for the asymptotically large values of $m$, the maximum of
the imaginary part of the quasi-normal frequency approaches zero
only asymptotically. The corresponding value of the rotation
parameter depends linearly on $m$ and, therefore, also approaches
infinity (see the right figure \ref{fig:D7_fit}). Thus, we
conclude that at least for the finite rotation parameter all modes
of negative $m$ are stable.

In figures \ref{fig:D8}, \ref{fig:D9}, \ref{fig:D11},
\ref{fig:a7m1} the dependence of QNMs on $j$ is shown. There, one
can see that the QN spectrum stays well far from the instability
point. Thus, we conclude that no growing mode was found for simply
rotating Myers-Perry black holes with $D=7, 8, 9, 10$ and $11$,
i.e. tensor-type gravitational perturbations are stable in the
above cases.

\section{Conclusions}

We have investigated the stability of simply rotating Myers-Perry black holes with $D\geq7$ against tensor-type gravitational perturbations. We have not found a growing mode, which apparently signifies stability against tensor perturbations.

In addition, we gave here extensive numerical data for the
quasinormal frequencies for the tensor-type gravitational
perturbations of these black holes.

Let us note that the quasinormal spectrum of a simply rotating
Myers-Perry black hole is not a limit of that of a MP black hole
in the AdS spacetime, when the AdS radius $R$ approaches infinity.
From the equation (71) of \cite{Kodama4}, one can see, that when
$R\gg r_+$ the quasinormal modes approach zero as
\begin{eqnarray}\nonumber
\omega&\sim&\frac{2n + D + 2l + m + j - 1 + F(D, l, m, j,
a)}{R}=\\\nonumber &=&{\cal O}\left(\frac{1}{R}\right).
\end{eqnarray}

In the forthcoming paper \cite{Kanti:2009sn}, we shall consider
graviton emission in the bulk for simply rotating black holes as
well for black holes with all equal angular momenta.


\begin{acknowledgments}
H. K. was supported in part by Grants-in-Aid for Scientific
Research from JSPS (No. 18540265). At the beginning of this work
R. A. K. was supported by \emph{the Japan Society for the
Promotion of Science (Japan)} and at the final stage by \emph{the
Alexander von Humboldt Foundation (Germany)} R. A. K. also
acknowledges the hospitality of the Theory Division of the High
Energy Accelerator Research Organization (KEK) in Tsukuba, where a
part of this work was done. A. Z. was supported by
\emph{Funda\c{c}\~ao de Amparo
\`a Pesquisa do Estado de S\~ao Paulo (FAPESP)}, Brazil.
\end{acknowledgments}

\end{document}